\documentclass[aps,pra,reprint,amsmath,amsfonts,amssymb,superscriptaddress]{revtex4-1}
\usepackage{graphicx,float,calc}
\usepackage{color,bm}
\usepackage{ulem}
\usepackage{braket}
\usepackage[colorlinks,urlcolor=blue,citecolor=blue,linkcolor=blue]{hyperref}
\begin{document}
\title{Topological Anderson insulators with different bulk states in quasiperiodic chains}
\author{Ling-Zhi Tang}
\affiliation{Guangdong Provincial Key Laboratory of Quantum Engineering and Quantum Materials, School of Physics and Telecommunication Engineering, South China Normal University, Guangzhou 510006, China}
\author{Shu-Na Liu}
\affiliation{Guangdong Provincial Key Laboratory of Quantum Engineering and Quantum Materials, School of Physics and Telecommunication Engineering, South China Normal University, Guangzhou 510006, China}
\author{Guo-Qing Zhang} 
\affiliation{Guangdong Provincial Key Laboratory of Quantum Engineering and Quantum Materials, School of Physics and Telecommunication Engineering, South China Normal University, Guangzhou 510006, China}
\affiliation{Guangdong-Hong Kong Joint Laboratory of Quantum Matter, Frontier Research Institute for Physics, South China Normal University, Guangzhou 510006, China}

\author{Dan-Wei Zhang}\thanks{danweizhang@m.scnu.edu.cn}
\affiliation{Guangdong Provincial Key Laboratory of Quantum Engineering and Quantum Materials, School of Physics and Telecommunication Engineering, South China Normal University, Guangzhou 510006, China}
\affiliation{Guangdong-Hong Kong Joint Laboratory of Quantum Matter, Frontier Research Institute for Physics, South China Normal University, Guangzhou 510006, China}

\begin{abstract}

We investigate the topology and localization of one-dimensional Hermitian and non-Hermitian Su-Schrieffer-Heeger chains with quasiperiodic hopping modulations. In the Hermitian case, phase diagrams are obtained by numerically and analytically calculating various topological and localization characters. We show the presence of topological extended, intermediate, and localized phases due to the coexistence of independent topological and localization phase transitions driven by the quasiperiodic disorder. Unlike the gapless and localized TAI phase in one-dimensional random disordered systems, we uncover three types of quasiperiodic-disorder-induced gapped topological Anderson insulators (TAIs) with extended, intermediate (with mobility edges), and localized bulk states in this chiral chain. Moreover, we study the non-Hermitian effects on the TAIs by considering two kinds of non-Hermiticities from the non-conjugate complex hopping phase and asymmetric hopping strength, respectively. We demonstrate that three types of TAIs preserve under the non-Hermitian perturbations with some unique localization and topological properties, such as the non-Hermitian real-complex and localization transitions and their topological nature. Our work demonstrates that the disorder-induced TAIs in Hermitian and non-Hermitian quasiperiodic systems are not tied to Anderson transitions and have various localization properties.

\end{abstract}

\date{\today}

\maketitle

\section{Introduction}
Topological insulators, hosting topological invariants for bulk states and nontrivial in-gap edge modes, have been widely explored in condensed matter \cite{XLQi2011,Hasan2010} and artificial systems \cite{DWZhang2018,Cooper2019,Goldman2016,Schroer2014,Roushan2014,XTan2018,XTan2019b,Lee2018b,Huber2016,LLu2014,Ozawa2019}. Due to the global topology of band structures, topological insulators are immune to weak disorder or local perturbations. However, strong disorders usually drive the systems into trivial gapless insulators as all the bulk states becomes Anderson localized \cite{Anderson1958}. Unexpectedly, it was theoretically found that a topological phase transition from trivial phases to topological insulators with robust edge states can be driven by moderate disorders \cite{JLi2009}. Such a disorder-induced topological phase is dubbed as topological Anderson insulators (TAIs). Actually, the underlying mechanism of TAIs is the renormalization of topological terms by disorders \cite{JLi2009,Groth2009,HJiang2009,YYZhang2012,SQShen2013}, instead of the disorder-induced localization phenomenon.

In recent years, the TAIs and their generalizations have been revealed in various systems \cite{JLi2009,Groth2009,HJiang2009,YYZhang2012,SQShen2013,HMGuo2010,Altland2014,Mondragon-Shem2014,Titum2015,BLWu2016,Sriluckshmy2018,JHZheng2019,Kuno2019,RChen2019,XSWang2020,CALi2020,YBYang2021,Velury2021}, even in some non-Hermitian systems \cite{DWZhang2019,XWLuo2019,HWu2020,LZTang2020,HLiu2020,QLin2021,Jahan2021} and in the presence of inter-particle interactions \cite{GQZhang2021,KLi2021,TCLI2021}. Some of them have been experimentally observed in engineered lattices, such as cold atomic gases \cite{Meier2018}, photonic and sonic crystals \cite{Stutzer2018,GGLiu2020,ZangenehNejad2020}, electric circuits \cite{WXZhang2021}, and photonic quantum walks \cite{QLin2021}. However, the interplay between disorder-induced topological and localization transitions remains largely unexplored. In most of these work, random disorders are considered and thus bulk states of the TAIs are fully localized. In particular, the TAI phase in one-dimensional (1D) Su-Schrieffer-Heeger (SSH) model \cite{WPSu1979} induced by random hopping strengths is gapless and contains only localized bulk states \cite{Altland2014,Mondragon-Shem2014,Meier2018}, without mobility edges and localization transition as the common wisdom that all states are Anderson localized in 1D random disordered systems \cite{Anderson1958}.

Quasiperiodic systems with incommensurate modulations in the potential or hopping terms are an ideal platform to study the Anderson localization and topological phases of matter \cite{Harper1955,Aubry1980,Thouless1983,Roati2008a,Luschen2018,Lahini2009,Kraus2012a,Shankar2013,Verbin2013,Madsen2013,LJLang2012,Ganeshan2013,Satija2013,Kraus2012,Tezuka2012,DeGottardi2013,HJiang2019,TLiu2020b,LJZhai2020,HJiang2019,TLiu2021,TXiao2021,Nakajima2021}. The quasiperiodic disorder can lead to localization phenomena without counterparts of random disorder in low dimensions, such as the localization transition \cite{Harper1955,Aubry1980,Thouless1983,Roati2008a}, the intermediate phase consisting of both localized and extended states \cite{YXLiu2020,CMDai2018,RoyShilpi2021,YTHsu2018,Luschen2018,SLXu2019,XPLi2016}, and the critical phase consisting of only critically localized states \cite{Chang1997,FLLiu2014,YCWang2021,YCWang2020,LZTang2021,TXiao2021}. Meanwhile, the topological charge pumping \cite{Thouless1983} can be realized in the paradigmatic 1D Aubry-Andr\'{e}-Harper model \cite{Harper1955,Aubry1980} and its variety of generalizations \cite{Kraus2012,Ganeshan2013,Satija2013,Kraus2012a,Shankar2013,Verbin2013,Madsen2013,LJLang2012,Ganeshan2013,Ganeshan2013}.
Very recently, the topological phase with critically localized bulk states in a quasiperiodic lattice \cite{TXiao2021} and the non-quantized pumping induced by the quasiperiodic disorder \cite{Nakajima2021} were experimentally observed with cold atoms. However, it remains unclear whether the quasiperiodic disorder can induce TAIs with different localization properties of bulk states. This question is essentially important for studying the mechanism and gap nature of TAIs \cite{JLi2009,Groth2009,HJiang2009,YYZhang2012,SQShen2013}, such as the bulk gap or mobility gap (mobility edges) for this disorder-induced topological phase. It is also practically interesting in experiments since the localization properties of bulk states are related to non-equilibrium dynamics or transport and the topology of TAIs can be measured from bulk dynamics \cite{Meier2018}.

On the other hand, growing effort has recently been made to explore many exotic topological and localization phenomena unique to non-Hermitian Hamiltonians or systems \cite{El-Ganainy2018,Ashida2020,Bergholtz2021,ZGong2018,Lee2016,SYao2018,Kunst2018,Zhang2020,Yang2020,Okuma2020,Borgnia2020,FSong2019,Hatano1996,Hatano1997,QBZeng2020,XMCai2021,HJiang2019,QBZeng2020,YXLiu2020,Longhi2019b,LWZhou2021,XMCai2021,Longhi2019a,CWu2021},
such as exceptional physics, new topological invariants and bulk-boundary correspondence, non-Hermitian skin effect, non-Hermitian delocalization and generalized mobility edges. The non-Hermiticity can be usually experimentally realized by gain and (or) loss \cite{El-Ganainy2018}. In very recent experiments, another two kinds of non-Hermiticities from effective non-reciprocal hoppings \cite{Hatano1996,Hatano1997} and complex on-site potentials \cite{Longhi2019a} have been engineered in ultracold atoms \cite{WGou2020}, photonic systems \cite{Weidemann2022,LXiao2020,QLin2021} and electrical circuit \cite{Helbig2020}. In the presence of non-reciprocal hopping or complex on-site potential, it has been revealed that topological phase transition characterized by a spectral winding number coincides with localization transition and (or) real-complex transition \cite{Longhi2019a,HJiang2019,YXLiu2020,LZTang2021,CWu2021}. Remarkable, exotic non-Hermitian TAIs induced by non-reciprocal hopping terms with random disorders in the generalized SSH model has been proposed and observed \cite{DWZhang2019,XWLuo2019,HWu2020,LZTang2020,HLiu2020,QLin2021}. It would be interesting to search for the disorder-induced TAIs in non-Hermitian quasiperiodic systems, especially in the presence of non-reciprocal hoppings and complex on-site potentials that have been recently realized.

In this work, we explore the interplay of topology and localization in Hermitian and non-Hermitian SSH chains with quasiperiodic hopping disorders, which are already realizable in some artificial systems \cite{Meier2018,TXiao2021,Helbig2020,LXiao2020,QLin2021,Weidemann2022,LXiao2020}. In the Hermitian case, we obtain the phase diagrams by numerically calculating various topological and localization properties. The numerical results consist with the analysis of topological boundaries obtained from the localization length of zero modes and the self-consistent Born approximation (SCBA). We show the topological extended, intermediate (partially localized), and localized phases due to the coexistence of topological and localization phase transitions driven by the quasiperiodic disorder. In particular, we uncover three types of gapped disorder-induced TAIs with extended, intermediate (consisting of mobility edges for extended and localized eigenstates), and localized bulk states in this system. These gapped TAIs driven from quasiperiodic disorders are different from the gapless and fully localized TAI in the SSH model with random disordered hoppings \cite{Altland2014,Mondragon-Shem2014,Meier2018}. The TAIs with different localization properties can be distinguished from the bulk dynamics or transport. Moreover, we study the non-Hermitian effects on the TAIs by considering two kinds of non-Hermiticities from the non-conjugate complex hopping phase and asymmetric hopping strength, respectively. We find that the proposed three types of TAIs preserve under the non-Hermitian perturbations and reveal some unique localization and topological properties in these two cases, such as the non-Hermitian real-complex and localization transitions and their topological nature. Thus, our work demonstrates that the disorder-induced TAIs in Hermitian and non-Hermitian quasiperiodic systems are not tied to Anderson transitions and have various localization properties, such as gapped TAIs with mobility edges.

The rest of the paper are organized as follow. We first reveal three types of TAIs in the Hermitian SSH chain with the quasiperiodic hopping modulation in Sec.\ref{sec2}. Section \ref{sec3} is then devoted to investigate non-Hermitian effects on the topological and localization properties of the uncovered TAIs. A brief conclusion is presented in Sec.\ref{sec4}.

\section{\label{sec2}TAIs with different bulk states}

We start by considering a generalized SSH model in a 1D dimerized lattice (denoted by $A$ and $B$ sublattices) with quasiperiodic disordered hopping. The system is described by the following tight-binding Hamiltonian
\begin{equation}\label{H}
	H=\sum_{n=1}^{N} (m_n a_n^\dagger b_n + t a_{n+1}^\dagger b_n + \text{H.c.}),
\end{equation}
where $N$ is the number of unit cell, $a_n^\dagger$ ($b_n$) denotes the creation (annihilation) operator for a particle on the $A$ ($B$) sublattice of the $n$-th cell, and $t$ and $m_n$ is the constant inter-cell hopping strength and the site-dependent intra-cell hopping strength, respectively. In this Hermitian system, we consider the quasiperiodic modulation on the intra-cell hopping term as
\begin{equation}\label{mn}
	m_n=m+W\cos(2\pi \alpha n + \theta),
\end{equation}
where $m$ is an overall intra-cell hopping strength, $W$ denotes the quasiperiodic disorder, $\theta$ is the additional phase shift and $\alpha$ is chosen as an irrational number to ensure the incommensurate modulation. In the clean limit with $W=0$, the Hamiltonian in Eq. (\ref{H}) reduces to the original SSH model with topological (trivial) phase when $m<t$ $(m>t)$, which is characterized by the 1D winding number and protected by the chiral (sublattice) symmetry. Notably, the generalized quasiperiodic SSH model described by Eq. (\ref{H}) can be realized by using cold atoms in a momentum-space lattice with tunable hopping modulations \cite{Meier2018,TXiao2021}.

In the following, we investigate the topology and localization in the model with the quasiperiodic hopping disorder, which still preserves the chiral symmetry. We set $t=1$ as the energy unit, $\alpha=(\sqrt{5}-1)/2$ as the golden ratio, the phase shift $\theta=0$, and the system size $N=L/2=610$ with $L$ being the total lattice number. Note that one can takes ensemble of the phase shift to analyze the localization properties in quasiperodic systems. However, we have confirmed that the topological and localization phase diagrams obtained for $\theta=0$ and $N=610$ in our numerical simulations preserve for other values of $\theta$ and larger system sizes (see Fig. \ref{fig7} for example). This is due to the fact that such a system size is large enough for self-averaging and neglecting finite-size effects. In certain case, we will take finite-size scaling analysis [see Fig. \ref{fig2}(g)]. The periodic boundary conditions (PBCs) is considered unless mentioned otherwise.

\begin{figure}[tb]
\centering
\includegraphics[width=0.48\textwidth]{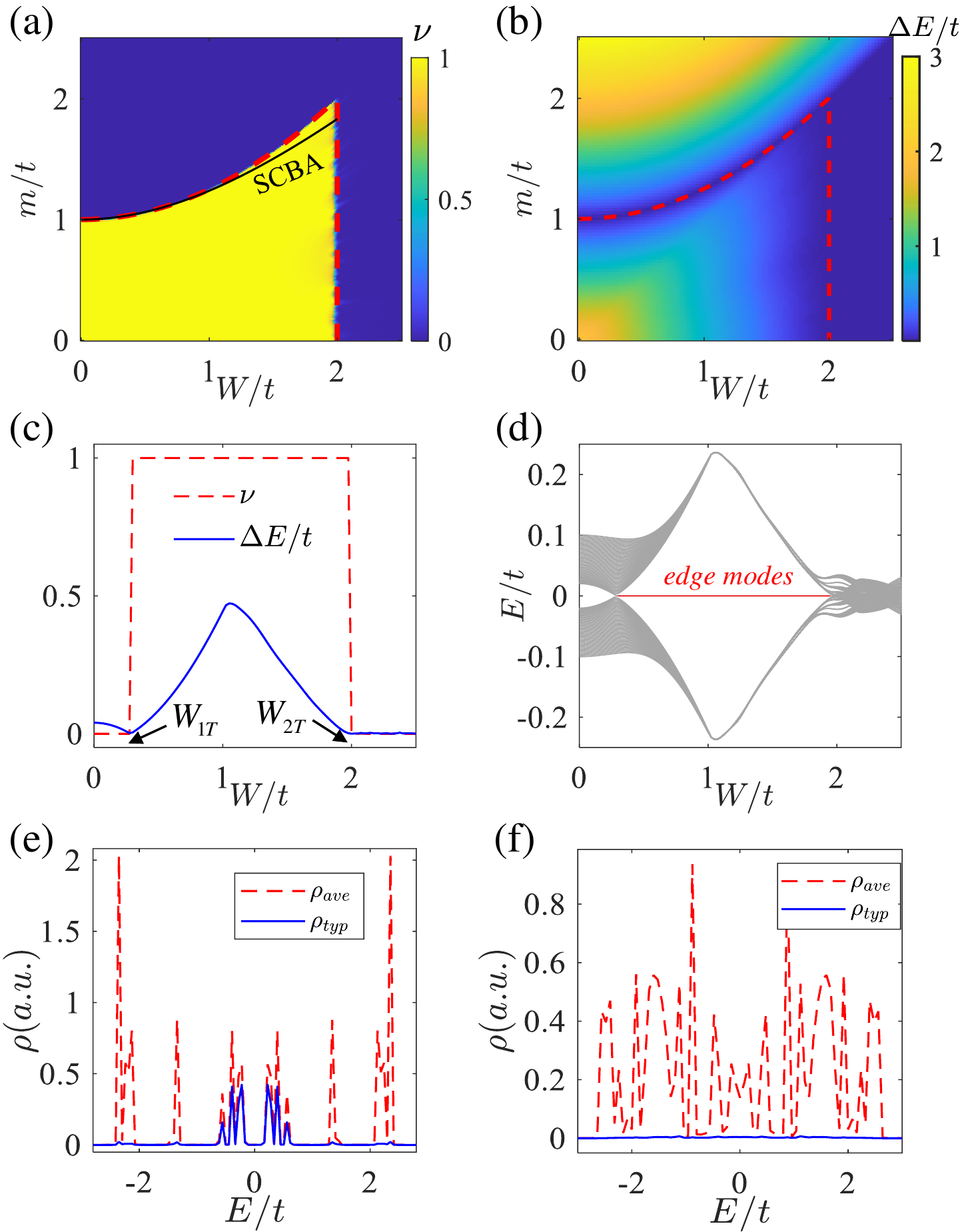}
\caption{(Color online) (a) Real-space winding number $\nu$, (b) energy gap $\Delta E$ as functions of $W$ and $m$. The red dash and black solid lines denote the topological phase boundaries determined by the divergence of the localization length of zero-energy states and from the SCBA analysis, respectively. (c) $\nu$ (red dash line) and $\Delta E$ (blue solid line) as a function of $W$ for $m=1.02$ with $W_{1T}$ and $W_{2T}$ being the topological transition points. (d) The middle 100 eigenenergies as a function of $W$ for $m=1.02$ under OBCs, with two zero-energy edge modes when $W_{1T}<W<W_{2T}$ colored in red. (e, f) Average DOS $\rho_{ave}$ (red dash line), typical DOS $\rho_{typ}$ (blue dot line) and $v$ (black solid line) as a function of Fermi energy for (e) quasiperiodic and (f) random disorder cases (averaged over 100 samples) with the same parameters $m=1.02$, $W=0.9$, and $N=610$.}\label{fig1}
\end{figure}

\subsection{Topological phase diagram and transitions}
To characterize the topological properties of the disordered chiral chain that breaks the translation symmetry, we use the real-space winding number defined by \cite{Mondragon-Shem2014}
\begin{equation}\label{winding}
	\nu= \frac{1}{L^\prime} \text{Tr}^\prime(\Gamma Q [Q,X]).
\end{equation}
Here $Q=\sum_{j=1}^N (\ket{j}\bra{j})-\ket{\tilde{j}}\bra{\tilde{j}}$ is obtained by solving the equation $H\ket{j}=E_{j}\ket{j}$ and $\ket{\tilde{j}}=\Gamma^{-1} \ket{j}$ with eigenenergies $E_{j}$ and eigenstates $\ket{j}$, $\Gamma=I_N \bigotimes \sigma_z$ is the chiral symmetry operator with the identity matrix $I_N$ and the Pauli matrix $\sigma_z$, $X$ is the coordinate operator, and $\text{Tr}^\prime$ denotes the trace over the middle interval of the lattice with the length $L^\prime=L/2$.

The topological phase diagram on the $W$-$m$ plane obtained by numerically computing $\nu$ is shown in Fig. \ref{fig1}(a). In the clean limit $W=0$, the topological transition between the trivial phase with $\nu=0$ and the topological phase with $\nu=1$ occurs at $m=1$.
The topological region for the constant intra-cell hopping strength $m$ enlarges when increasing $W$ to a modest regime. One can find the TAIs driven by the quasiperiodic disorder from the trivial phase when $1<m\lesssim2$. The disorder-induced TAI regime is similar to that in the SSH model with random hopping disorders \cite{Altland2014,Mondragon-Shem2014,Meier2018}, however, it contains three different kinds of bulk states as we will discuss below.

We also compute the bulk gap $\Delta E = E_{N+1}-E_{N}$ under the PBCs, as shown in Fig. \ref{fig1}(b). The bulk gap closes at the topological transition points and becomes vanishing for trivial gapless Anderson insulators when $W$ is large enough ($W\gtrsim2$). To be more clear, we plot $\nu$ and $\Delta E$ as a function of $W$ for fixed $m=1.02$ in Fig. \ref{fig1}(c). One can find the disorder-induced TAIs lying between the first and second topological transition points at $W_{1T}\approx0.27$ and $W_{2T}\approx2.0$, respectively. The disorder-induced zero-energy edge modes in the energy spectrum of the TAIs phase under the open boundary conditions (OBCs) are shown in Fig. \ref{fig1}(d), owing to the bulk-boundary correspondence.

At the topological transition points, the localization length of zero-energy modes are divergent due to their delocalization nature in 1D chiral chains \cite{Mondragon-Shem2014}.
For our model Hamiltonian, the wave function of zero-energy eigenstate $\psi=\{\psi_{1,A},\psi_{1,B},\psi_{2,A},\psi_{2,B}\cdots \psi_{N,A},\psi_{N,B}\}^T$ can be obtained by solving the corresponding Schr\"{o}dinger equation $H\psi=0$. This corresponds to the eigen-equations
$t\psi_{n,B}+m_n \psi_{n+1,B}=0$ and $m_n\psi_{n,A}+t \psi_{n+1,A}=0$, which leads to the form of the probability distribution of the zero-energy wave function
\begin{equation}
\begin{aligned}
	\psi_{n, A}&=(-1)^{n} \prod_{l=1}^{n}\frac{m_l}{t} \psi_{1, A}, \\
\psi_{n, B}&=(-1)^{n} \prod_{l=1}^{n}\frac{t}{m_{l+1}} \psi_{1, B}.
\end{aligned}
\end{equation}
Thus, the inverse of localization length ($\Lambda$) of zero-energy modes in the limit $N\to \infty$ reads
\begin{equation}\label{Lambda-}
	\Lambda^{-1}=\max \left\{\lim _{N \rightarrow \infty} \frac{1}{N} \ln \left|\psi_{N, A}\right|, \lim _{N \rightarrow \infty} \frac{1}{N} \ln \left|\psi_{N, B}\right|\right\}.
\end{equation}
By setting $\psi_{1,A}=\psi_{1,B}=1$, one can obtain
\begin{equation}\label{Lambda--}
	\begin{aligned} \lim _{N \rightarrow \infty} \frac{1}{N} \ln \left|\psi_{N, A}\right| &=\lim _{N \rightarrow \infty} \frac{1}{N} \ln \left|\psi_{N, B}\right| \\ &=\mid \lim _{N \rightarrow \infty} \frac{1}{N} \sum_{l=1}^{N}\left(\ln \left|t\right|-\ln \mid m_{l}|\right)|. \end{aligned}
\end{equation}
Substituting Eq. (\ref{Lambda-}) into Eq. (\ref{Lambda--}), we obtain
\begin{equation}\label{loclength}
	\Lambda^{-1}=\mid \lim _{N \rightarrow \infty} \frac{1}{N} \sum_{l=1}^{N}\left(\ln \left|t\right|-\ln \mid m_{l}|\right)|.
\end{equation}
Note that $\Lambda^{-1}\rightarrow0$ when the localization length diverges ($\Lambda\rightarrow\infty$). We show the results of $\Lambda^{-1}\approx0$ [numerically obtained by solving Eq. (\ref{loclength})] as the red dash lines in Figs. \ref{fig1}(a) and \ref{fig1}(b). The results indicate that topological and trivial phases can be well separated by the delocalization nature of zero modes in this quasiperiodic system.

\begin{figure*}[tb]
\centering
\includegraphics[width=1\textwidth]{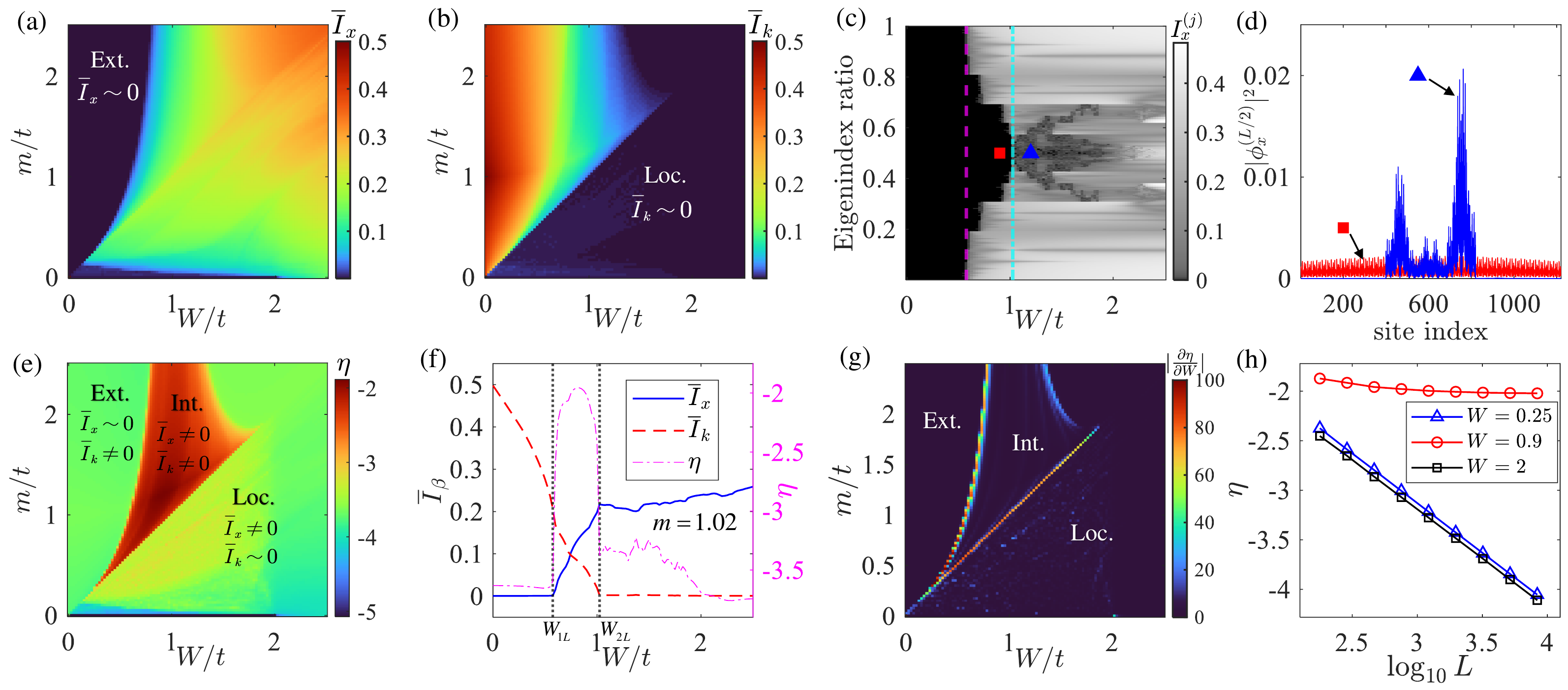}
\caption{(Color online) Mean IPRs $\bar{I}_x$ (a) and $\bar{I}_k$ (b) and quantity $\eta$ (e) on the $W$-$m$ plane. (c) The eigenstate real-space IPR $I_{x}^{(j)}$ as a function of $W$ for fixed $m=1.02$. (d) Density distribution of $L/2$-th (middle) eigenstates $|\phi_{x}^{(L/2)}|^2$  for $W=0.9$ (an extended state) and $W=1.2$ (a localized state) labeled in (c). (f) $\bar{I}_x$ (blue solid line), $\bar{I}_k$ (red dash line) and $\eta$ (pink dash-dot line) as a function of $W$ for $m=1.02$. (g) The derivation quantity $|\partial\eta/\partial W|$ for determining the boundaries between three localization phases on the $W$-$m$ plane. (h) Scaling of $\eta$ with lattice size $L$ for three localization phases with different values of $W$ and $m=1.02$. In (a,b,e,g), the regions for extended, localized and intermediate phases are denoted by 'Ext.', 'Loc.' and 'Int.', respectively. In (c,f), the vertical lines with $W_{1L}\approx0.57$ and $W_{2L}\approx1.02$ denote the first and second localization transitions, respectively.}\label{fig2}
\end{figure*}

We further perform the SCBA analysis to reveal the disorder-induced renormalization of the topological term for the TAIs in the topological phase diagram [Fig. \ref{fig1}(a)]. This analysis works in the region $W\lesssim2$ where the disorder is not dominated. Based on the effective medium theory and the SCBA method \cite{Groth2009}, one can self-consistently obtain the disorder-induced self-energy as the renormalization of a clean Hamiltonian. For the Hamiltonian in Eq. (\ref{H}), the self-energy term $\Sigma(W)$ satisfies the self-consistent equation
\begin{equation}
	\frac{1}{E_F-H_q(k)-\Sigma(W)}=\braket{\frac{1}{E_F-H_{\text{eff}}(k,W)}},
\end{equation}
where $E_F\equiv0$ is Fermi energy, $H_q(k)=[m+t\cos(k)]\sigma_x+t\sin(k)\sigma_y$ is the clean Hamiltonian ($W=0$) in momentum space with $\sigma_{x,y}$ the Pauli matrices, $\Sigma=\Sigma_x \sigma_x + \Sigma_y \sigma_y$ and the $\braket{\cdots}$ denotes averaging over all disorder samples. In our model, the quasiperiodic disorder follows the form $V(n)\sigma_x$ with $V(n)=W\cos(2\pi\alpha n)$, thus the effective Hamiltonian $H_{\text{eff}}=H_q(k)+V(n)\sigma_x$. Considering the symmetry of the Hamiltonian, the self-energy is simplified as $\Sigma(W)=\Sigma_x(W) \sigma_x$. The intra-cell hopping strength $m$ is then renormalized as $\bar{m}=m+\Sigma_x(W)$. The topological phase boundary on the $m$-$W$ plane is determined by $\bar{m}(m,W)=t$. The numerical results of the topological phase boundary based on the SCBA for small and modest disorder strength is shown as the black solid line in Fig. \ref{fig1}(a), which agree well with that determined by the winding number.

So far, we have shown that the origin for emergence of TAIs in this quasiperiodic chiral chain is the band inversion and the renormalization of topological terms driven by the disorder. This mechanism is the same as that for TAIs in random disordered systems \cite{JLi2009,Groth2009,HJiang2009,YYZhang2012,SQShen2013}. However, comparing with the gapless and fully localized TAI phase in the SSH model with random disordered hoppings, the TAI phase in this quasiperiodic case is gapped and can be partially localized (or fulled delocalized). We have numerically check that the finite bulk gap $\Delta E$ in the TAI phase shown in Figs. \ref{fig1}(b-d) preserve for larger lattice size $L$, while $\Delta E=0$ for the random disorder SSH chain \cite{Altland2014,Mondragon-Shem2014,Meier2018}. Following Refs. \cite{YYZhang2012,SQShen2013}, we can numerically compute local density of states (DOS)
\begin{equation}
 	\rho(l,E)=\frac{1}{L}\sum_{j=1}^{L}|\braket{l|j}|^2\delta(E-E_j)
\end{equation}
and the arithmetic mean of the local DOS
\begin{equation}
    \rho_{ave}(E)=\langle\langle \rho(l,E) \rangle\rangle,
\end{equation}
where $\langle\langle \cdots \rangle\rangle$ denotes the average over the site $l$ of the lattice. Here $\rho_{ave}(E)$ denote the bulk DOS, and $\rho_{ave}(E)=0$ corresponds to the gapped region for the Fermi energy $E_F=E$ under PBCs \cite{YYZhang2012,SQShen2013}. As shown in Fig. \ref{fig1}(e), we find $\rho_{ave}(E)=0$ for $-0.11\lesssim E\lesssim0.11$, corresponding the gapped TAI (for $m=1.02$ and $W=0.9$) with a bulk gap $\Delta E\approx0.22$. For comparison, in Fig. \ref{fig1}(f), we compute $\rho_{ave}(E)$ for the SSH model with uniform random disorder $m_n=m+\zeta_n$ and $\zeta_n \in [-W,W]$. We find finite DOS $\rho_{ave}(E)\neq0$ in the region corresponding to the gapless TAI induced by random disorders \cite{Meier2018,Mondragon-Shem2014}. Moreover, we compute the geometric mean of the local DOS \cite{YYZhang2012,SQShen2013}
\begin{equation}
    \rho_{typ}(E)=\exp[\langle\langle \ln \rho(l,E) \rangle\rangle],
\end{equation}
which characterizes localized and extended states around $E$ as $\rho_{typ}(E)/\rho_{ave}(E)\to0$ and $\rho_{typ}(E)/\rho_{ave}(E)\neq0$ in the large $L$ limit, respectively. Figure \ref{fig1}(e) shows the coexistence of extended and localized states, which indicates the emergence of mobility edges for the TAI in this quasiperiodic SSH chain (see the following section). For the corresponding TAI phase in the random disordered SSH chain \cite{Meier2018,Mondragon-Shem2014}, all the bulk states are localized with
vanishing $\rho_{typ}$, as excepted and shown in Fig. \ref{fig1}(f).

\subsection{Localization properties of bulk states}
In this section, we further study the localization properties of bulk states in this quasiperiodic SSH chain. Following Refs. \cite{YXLiu2020,CMDai2018,RoyShilpi2021,YTHsu2018,SLXu2019,XPLi2016}, we numerically compute the inverse participation ratio (IPR) of the $j$-th eigenstate in real and momentum spaces
\begin{equation}
	I_{\beta}^{(j)}= \sum_{l=1}^{L}\left|\phi_{\beta}^{(j)}(l)\right|^{4}
\end{equation}
where $\beta=x,k$ respectively denote the sum over the real (lattice) and momentum sites, and $\phi_{\beta}^{(j)}(l)$ denotes the probability amplitude of the $j$-th normalized eigenstate at $l$-th site in the $\beta$ space. In our numerical simulations, we directly compute $I_{x}^{(j)}$ from the real-space Hamiltonian and then obtain $I_{k}^{(j)}$ by using the discrete Fourier transformation $\phi_{k}^{(j)}(l)=\frac{1}{L} \sum_{p=1}^{L} e^{-i 2 \pi p l /L} \phi_{x}^{(j)}(p)$. For an extended ($j$-th) eigenstate, $I_{x}^{(j)}\sim L^{-1}$ and $I_{k}^{(j)}\sim \mathcal{O}(1)$, while $I_{x}^{(j)}\sim \mathcal{O}(1)$ and $I_{k}^{(j)}\sim L^{-1}$ for a localized eigenstate.

In order to define different localization phases, one can use the mean IPR averaged over all eigenstates
\begin{equation}
	\bar{I}_\beta= \frac{1}{L} \sum_{j=1}^{L} I_{\beta}^{(j)}.
\end{equation}
For the extended (localized) phase with all eigenstates being extended (localized) in the large $L$ limit, one has $\bar{I}_x\sim L^{-1}\sim0$ and $\bar{I}_k\sim\mathcal{O}(1)\neq0$ ($\bar{I}_k\sim L^{-1}\sim0$ and $\bar{I}_x\sim\mathcal{O}(1)\neq0$). Furthermore, one can define the intermediate phase if both localized and extended eigenstates exist in the energy spectrum with mobility edges (structures) \cite{YXLiu2020,CMDai2018,RoyShilpi2021,YTHsu2018,SLXu2019,XPLi2016}. Thus, both mean IPRs $\bar{I}_x$ and $\bar{I}_k$ are finite in the intermediate phase.

Figures \ref{fig2}(a) and \ref{fig2}(b) show the numerical results of $\bar{I}_x$ and $\bar{I}_k$ on the $W$-$m$ plane with $N=L/2=610$, respectively. We can find the extended and localized phase regions with $\bar{I}_x\sim0$ and $\bar{I}_k\sim0$, respectively. In addition, there is an intermediate phase region with $\bar{I}_{x,k}\neq0$ lying between the extended and localized phases on the $W$-$m$ plane. To figure out the localization properties in this region, we display the IPRs of all eigenstates $I_{x}^{j}$ as a function of $W$ for $m=1.02$ in Fig. \ref{fig2}(c). Such an intermediate phase consists of mobility edges for extended and localized eigenstates, which are lying between the first and second localization transition points at $W=W_{1L}\approx0.57$ and $W=W_{2L}\approx1.02$ (see also Fig. \ref{fig2}(f)]). Note that all the egienstates become localized when $W>W_{2L}$. For instance, we plot the real-space density distribution of the $L/2$-th (middle) eigenstate for $W=0.9$ and $W=1.2$ in Fig. \ref{fig2}(d), which are extended and localized, respectively.

The structure of mobility edge shown with respect to the disorder strength $W$ [see Fig. \ref{fig2}(c)] reveals the existence of the intermediate phase. In order to identify the intermediate (and other) phase region on the $W$-$m$ plane, we can define the dimensional quantity \cite{Xli2020}
\begin{equation}\label{eta}
	\eta=\log_{10}(\bar{I}_x \times \bar{I}_k).
\end{equation}
In the localized and extended phase region, either of $\bar{I}_x$ and $\bar{I}_k$ is $\sim\mathcal{O}(L^{-1})$, one has the quantity $\eta<-\log_{10}L$ and $\eta\propto\log_{10}L$, such as $\eta<-3$ for $L=2N=1220$ in our simulations. In contrast, in the intermediate phase region where both $\bar{I}_x$ and $\bar{I}_k$ are finite [$\sim\mathcal{O}(1)$], we obtain larger values of $\eta$ with $-2.5\lesssim\eta\lesssim-1.8$, as shown in Fig. \ref{fig2}(e). In this way, the quantity $\eta$ can be used to clearly identify the intermediate phase in the localization phase diagram. To be more clearly, Fig. \ref{fig2}(f) shows $\bar{I}_{x,k}$ and $\eta$ as a function of $W$ with fixed $m=1.02$, as an example. In the extended phase when $0\leqslant W<W_{1L}$ and the localized phase when $W>W_{2L}$ with $W_{1L}\approx0.57$ and $W_{2L}\approx1.02$, one has $\bar{I}_x\sim0$ and $\bar{I}_k\sim0$, respectively. In the intermediate phase when $W_{1L}<W<W_{2L}$, both $\bar{I}_x$ and $\bar{I}_k$ are finite. Notably, the quantity $\eta$ change sharply with respect to $W$ between three localization phases in Fig. \ref{fig2}(f). Thus, we can define the derivation quantity $|\partial\eta/\partial W|$ to further determine the phase boundaries, as shown in Fig. \ref{fig2}(g). Furthermore, we show the scaling of $\eta$ with the lattice size $L$ for three different localization phases in Fig. \ref{fig2}(g). As excepted, $\eta$ is insensitive to $L$ and tends to a constant $\eta(L\rightarrow\infty)\sim-2$ in the intermediate phase with $W=0.9$ and $m=1.02$, while $\eta$ decreases linearly with $\log_{10}(L)$ for the other two phases. Note that we have checked that in the intermediate phase region, the structure of mobility edges always exhibits in the energy spectrum, similar as that in Fig. \ref{fig2}(c).

\begin{figure}[tb]
	\centering
	\includegraphics[width=0.48\textwidth]{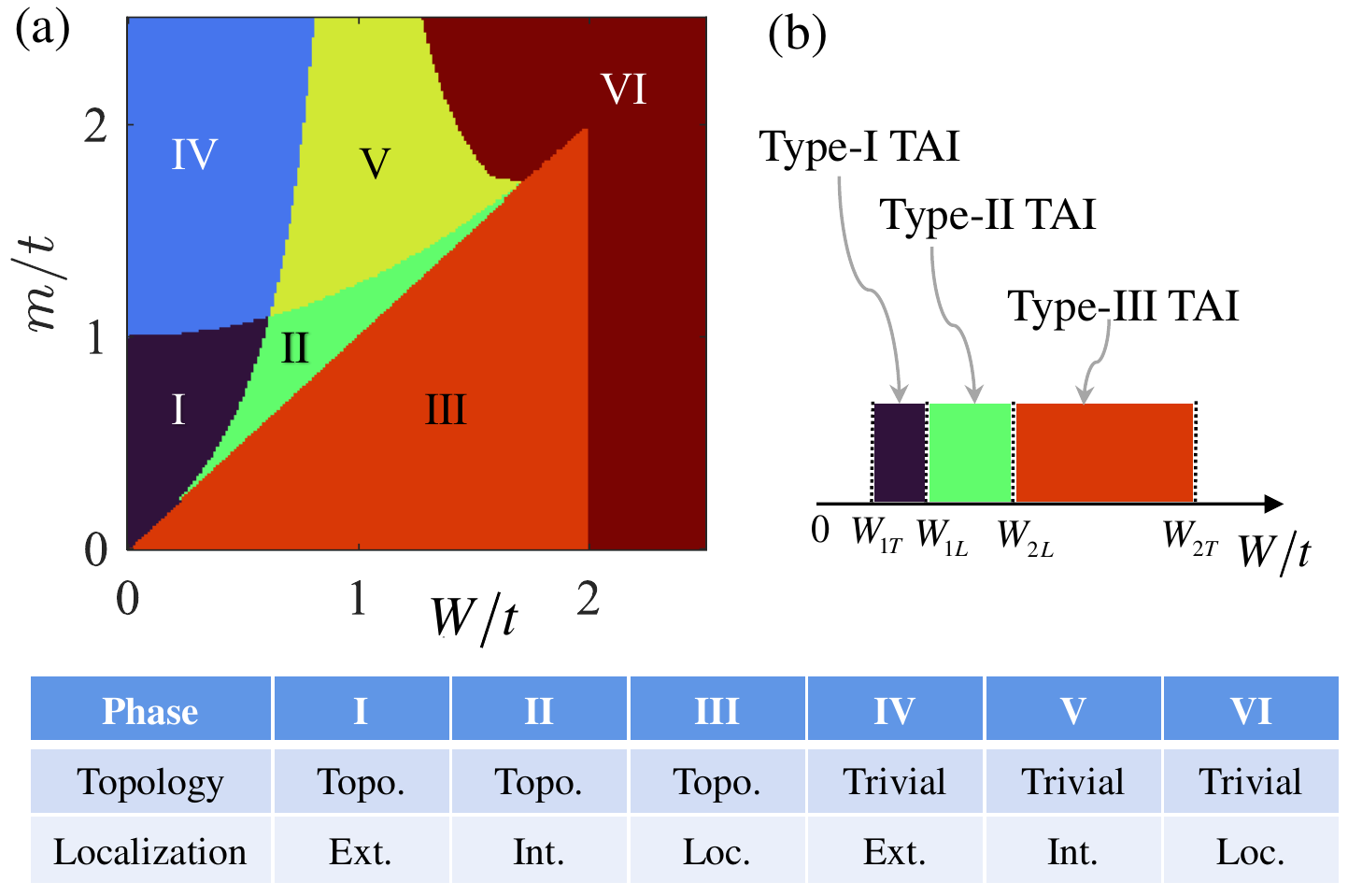}
	\caption{(Color online) (a) Total phase diagram on the $W$-$m$ space. There are six different phases: extended topological phase (I), topological intermediate phase (II), topological localized phase (III), trivial extended phase (IV), trivial intermediate phase (V), and trivial localized phase (VI), as illustrated in the table. (b) Regions of $W$ with fixed $m=1.02$ for three types of TAIs induced by the quasiperiodic disorder.
	}\label{fig3}
\end{figure}

Combining the topological and localization phase diagrams in Figs. \ref{fig1}(a) and \ref{fig2}(e,g), we finally obtain the total phase diagram on the $W$-$m$ plane, as shown in Fig. \ref{fig3}(a). There are six phases in the phase diagram determined by different topological and localization properties. They are labeled from I to VI: Phases I, II and III are topological with extended, intermediate, localized bulk states, respectively; and Phases IV, V and VI are trivial with extended, intermediate, localized bulk states, respectively. Thus, we obtain three types of TAIs with different localization properties of bulk states. For instance, as shown in Fig. \ref{fig3}(b) with $m=1.02$, one can find the disorder-driven transition from trivial extended phase to the TAIs with extended ($W_{1T}<W<W_{1L}$), intermediate ($W_{1L}<W<W_{2L}$) and localized bulk states ($W_{2L}<W<W_{2T}$), which are dubbed as type-I, type-II and type-III TAIs, respectively. For strong disorder ($W>W_{2T}$), the system becomes trivial gapless phase with fully localized states. Notably, the quasiperiodic disorder induced TAIs with extended and intermediate bulk states uncovered here are absent in random disordered systems. For instance, only the (type-III) TAIs with fully localized bulk states exhibits in the SSH model with random disorder hopping \cite{Altland2014,Mondragon-Shem2014,Meier2018}. 
Notably, different phases in the total phase diagram can be detected from the localization and topological properties. By measuring the bulk dynamics (transport) in the lattice \cite{Weidemann2022}, the extended, localized, and intermediate phases can be revealed from the ballistic, subdiffusive, and diffusive dynamics, respectively. The topological phase can be detected from the in-gap zero edge modes or from the topological winding number by also measuring the bulk dynamics \cite{Meier2018}.

\section{\label{sec3}Non-Hermitian TAIs}

In this section, we proceed to study the non-Hermitian effects on the three types of TAIs. In Sec. \ref{IIIA}, we consider the non-Hermiticity induced by non-conjugate complex hopping phase. We also consider the asymmetric hopping strength in Sec. \ref{IIIB}. We find that three types of TAIs preserve and exhibit some properties unique to the non-Hermitian systems.

\subsection{\label{IIIA}Non-conjugate hopping-phase case}
We introduce the non-Hermiticity to the hopping term of the model Hamiltonian in Eq. (\ref{H}), which now becomes
\begin{equation}\label{H'}
	H'=\sum_{n=1}^{N} m'_n a_n^\dagger b_n + m'_n b_n^\dagger a_n + (t a_{n+1}^\dagger b_n + \text{H.c.}),
\end{equation}
with the modified intra-cell hopping term
\begin{equation}\label{m'_n}
	m'_n=m+W\cos(2\pi \alpha n + \theta + ih).
\end{equation}
Here the non-Hermiticity comes from the complexification of the phase shift with an additional imaginary phase $h$ \cite{Longhi2019a}. It has been proposed that such a complex phase shift can be realized in non-Hermitian photonic quasicrystals coupled with axial cavity modes by a phase modulator \cite{Longhi2019a}. Very recently, the non-Hermitian photonic quasicrystal with the effective complex phase shift has been experimentally realized in Ref. \cite{Weidemann2022}. Note that the non-Hermitian Hamiltonian in Eq. (\ref{H'}) possesses the sublattice symmetry \cite{Kawabata2019b}, which is also refereed as the chiral symmetry in the Hermitian case.

\begin{figure}[tb]
	\centering
	\includegraphics[width=0.48\textwidth]{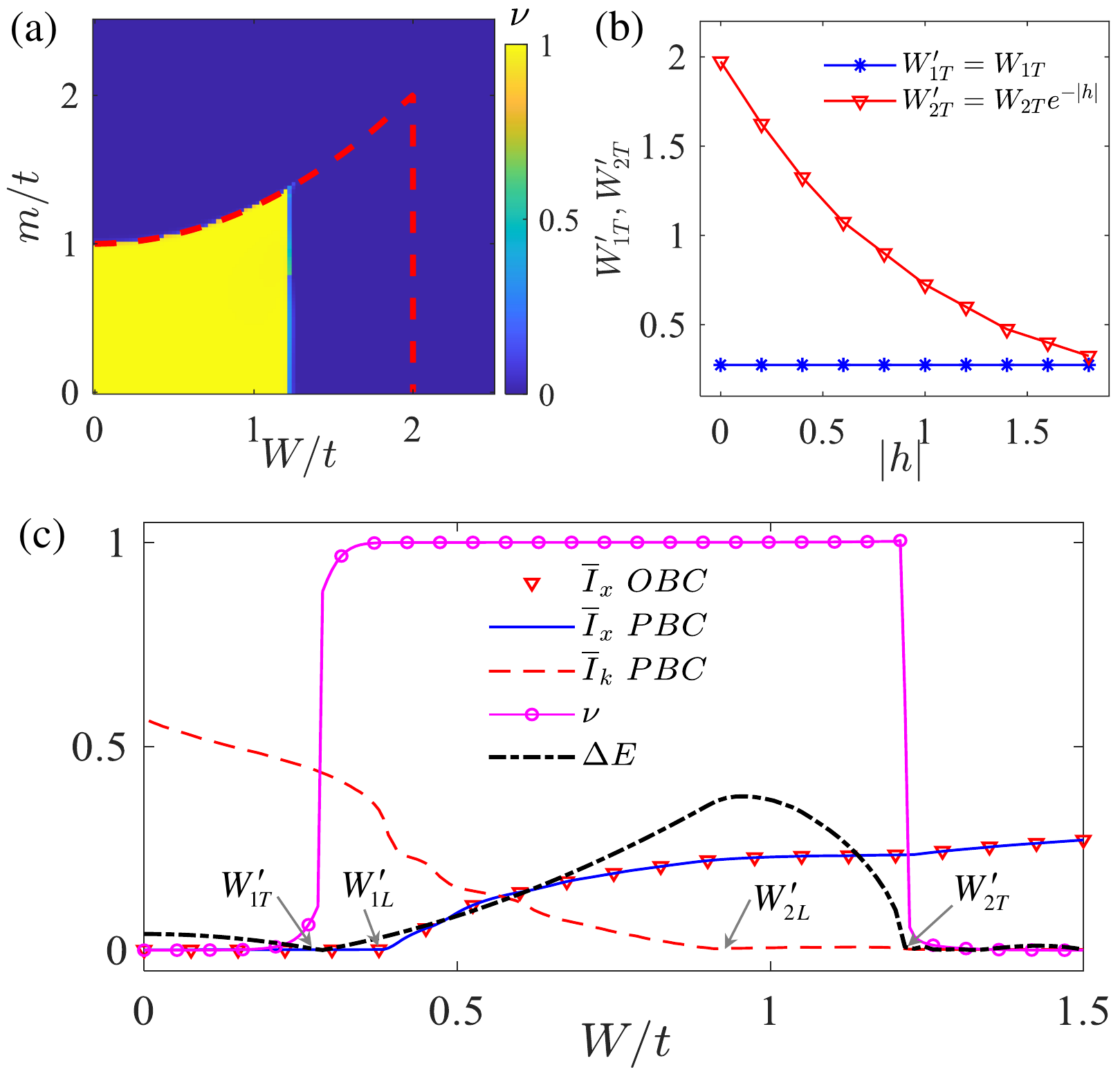}
	\caption{(Color online) (a) $\nu$ as functions of $W$ and $m$ for $h=0.5$. Red dash line denoted the topological phase boundary in the Hermitian case ($h=0$). (b) The first and second topological transition points $W'_{1T}$ and $W'_{2T}$ for varying non-Hermiticity parameter $|h|$ and fixed $m=1.02$. (c) $\nu$, $\bar{I}_{x,k}$ (under the PBCs and OBCs), and $\Delta E$ as a function of $W$ for $m=1.02$ and $h=0.5$. The two topological and localization transition points at $W'_{1T}\approx0.27$, $W'_{2T}\approx1.21$, $W'_{1L}\approx0.39$, and $W'_{2L}\approx0.92$ are labeled.
	}\label{fig4}
\end{figure}

As shown in Refs. \cite{FSong2019,DWZhang2019,XWLuo2019}, the real-space winding number $\nu$ can be generalized to non-Hermitian Hamiltonians under the biorthogonal basis. To this end, the matrix $Q$ in Eq. (\ref{winding}) is replaced by
$Q=\sum_{j=1}^N (\ket{j'R}\bra{j'L})-\ket{\tilde{j'}R}\bra{\tilde{j'}L}$.
Here right eigenstates $\ket{j'R}$ and left eigenstates $\ket{j'L}$ are obtained from the eigenfunctions $H'|j'R\rangle=E_{j'R}|j'R\rangle$ and $H^{\dagger}|j'L\rangle=E_{j' L}|j'L\rangle$ with $\ket{\tilde{j'}R}=\Gamma^{-1} \ket{j'R}$ and $\ket{\tilde{j'}L}=\Gamma^{-1} \ket{j'L}$, which form the biorthogonal basis. We numerically calculate $\nu$ on the $W$-$m$ plane for $h=0.5$, as shown in Fig. \ref{fig4}(a), where the red dash line denotes the topological phase boundary in the Hermitian case [$h=0$ in Fig. \ref{fig1}(a)]. One can find that this kind of non-Hermiticity with finite $|h|$ reduces the topological region by moving the second topological transition from $W_{2T}$ to $W'_{2T}<W_{2T}$, while keeps the first topological transition at $W'_{1T}=W_{1T}<W'_{2T}$. To be more clearly, we plot the $W'_{1T}$ and $W'_{2T}$ as a function of $|h|$ for $m=1.02$ in Fig. \ref{fig4}(b). They are well fitting by $W'_{1T}=W_{1T}$ and $W'_{2T}=W_{2T}e^{-|h|}$, with $W_{1T}\approx0.27$ and $W_{2T}\approx2.0$ in the Hermitian limit.

\begin{figure*}[tb]
	\centering
	\includegraphics[width=0.8\textwidth]{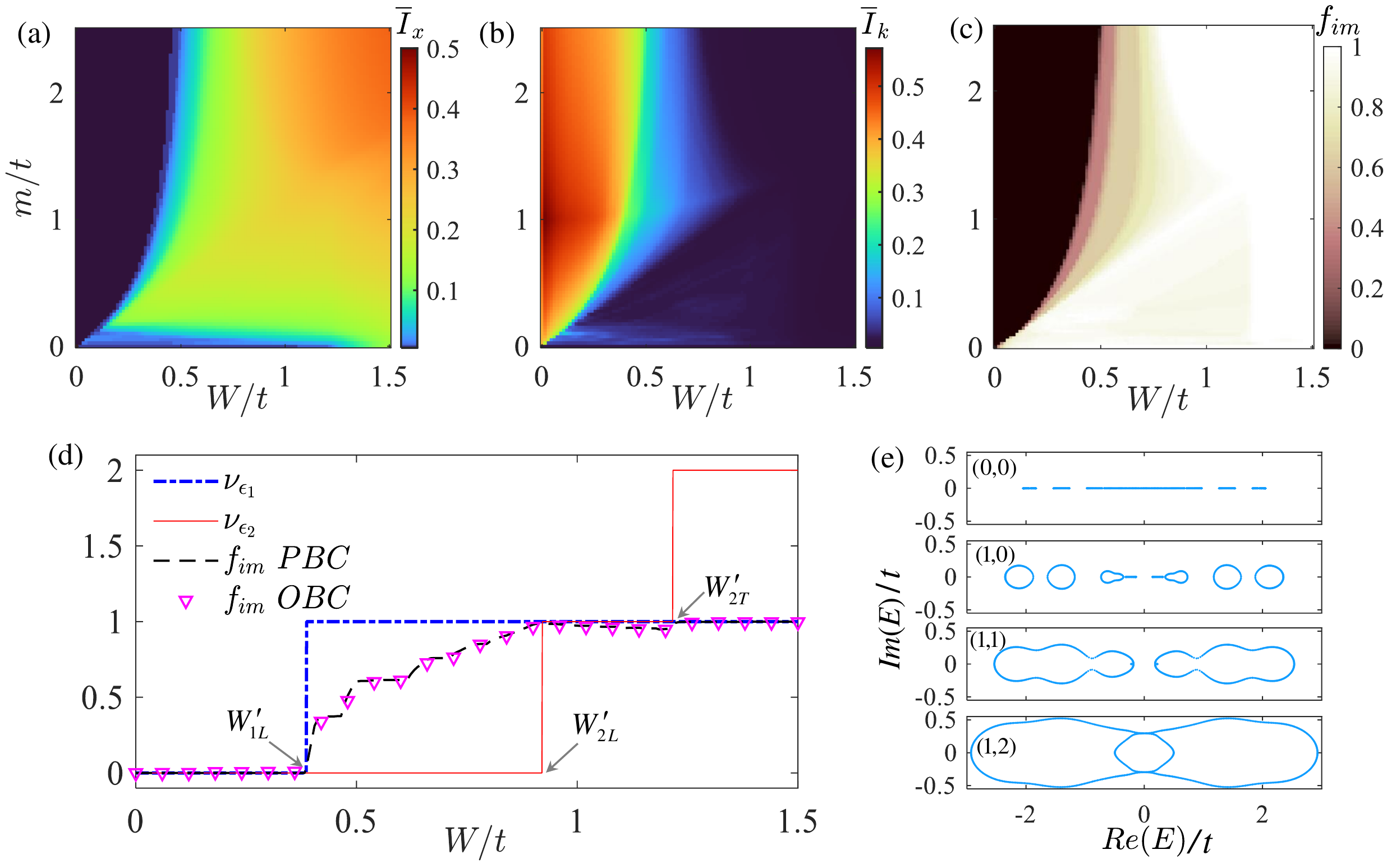}
	\caption{(Color online) Averaged IPR $\bar{I}_x$ (a) and $\bar{I}_k$ (b) and ratio of complex eigenenergies $f_{im}$ (c) under PBCs for $h=0.5$ on the parameter space $W$-$m$. (d) Plot of spectral winding number $\nu_{\epsilon_1}$ (blue dash line), $\nu_{\epsilon_2}$ (red dash line), $f_{im}$ for PBCs (black dash line) and $f_{im}$ for OBCs (pink inverted triangle) for $m=1.02,\ h=0.5$ as a function of $W$. The two localization transition points $W'_{1L}\approx0.39$ and $W'_{2L}\approx0.92$, and the topological transition point $W'_{2T}\approx1.21$ are labeled. (e) Typical energy spectrum for $W=0.3,0.8,1,1.4$ (from top to bottom) with the energy winding numbers $(\nu_{\epsilon_1},\nu_{\epsilon_2})$ in each panel for $h=0.5$. From top to bottom, the energy basis are $(\epsilon_1,\epsilon_2)$=$(0,2.05)$, $(0.13,2.35)$, $(1.85,2.54)$, $(0,2.93)$.}\label{fig5}
\end{figure*}

In Fig. \ref{fig4}(c), we plot the winding number $\nu$ and the energy gap $\Delta E=\text{Re}(E_{N+1,R}-E_{N,R})$ defined by the real part of the complex eigenenergies $E_{N,R}$ (sorted by the real part of spectrum) as a function of $W$ for $m=1.02$ and $h=0.5$. The two topological phase transitions happen at $W'_{1T}\approx0.27$ and $W'_{2T}\approx1.21$ with the gap closing. To reveal the localization properties in this non-Hermitian case, we calculate the IPRs $\bar{I}_{x,k}$ averaged over the right eigenstates $\ket{jR}$. The results of $\bar{I}_x$ are the same under PBCs and OBCs in both the clean and disordered cases. This indicates the absence of non-Hermitian skin effect \cite{SYao2018,Kunst2018,Zhang2020,Yang2020,Okuma2020,Borgnia2020} of bulk states under OBCs. With the results of $\bar{I}_\beta$, we can obtain the first and second localization transition points at $W'_{1L}\approx0.39$ and $W'_{2L}\approx0.92$. Thus, the three types of TAIs can still be induced by the quasiperiodic disorder in this non-Hermitian system when $0.27\lesssim W\lesssim0.39$ (type-I), $0.39\lesssim W\lesssim0.92$ (type-II), and $0.92\lesssim W\lesssim1.21$ (type-III), respectively. In Figs. \ref{fig5}(a) and \ref{fig5}(b), we further show the results of $\bar{I}_{x,k}$ under PBCs on the $W$-$m$ plane with $h=0.5$. Comparing to the results for $h=0$ in Figs. \ref{fig2}(a) and \ref{fig2}(b), we can find that the non-Hermiticity enlarges the parameter region of the localized phase and reduces the extended and intermediate phase regions. The result indicates that the non-conjugate hopping tends to localize the bulk states.

The real-complex transition of the energy spectrum and its winding on the complex energy plane are unique to non-Hermitian Hamiltonians \cite{El-Ganainy2018,Ashida2020,Bergholtz2021}. To study the real-complex transition, we numerically compute the ratio of the right eigenstates $\ket{jR}$ with complex eigenenergies in the energy spectrum, which is defined by
\begin{equation}
	f_{im}=L_{im}/L,
\end{equation}
with $L_{im}$ the number of eigenenergies whose imaginary part $|\text{Im}(E_{jR})|>10^{-13}$ as the cutoff in our simulations. The numerical result of $f_{im}$ for $h=0.5$ on the $W$-$m$ plane is shown in Fig. \ref{fig5}(c). We find that the energy spectrum is either real ($f_{im}=0$) or complex ($0<f_{im}\leqslant1$) in the phase diagram. The boundary between the real and complex energies corresponds to the localization-delocalization phase boundary in Fig. \ref{fig5}(a). To see the coincidence more clearly, we plot $f_{im}$ under PBCs and OBCs as a function of $W$ in Fig. \ref{fig5}(d), with the localization transition points $W'_{1L}$ and $W'_{2L}$ being labeled. One can see that $f_{im}$ turns to non-zero at $W'_{1L}$, which shows that the real-complex transition coincides with the localization transition from the extended to intermediate phases. In addition, $f_{im}\thickapprox1$ in the fully localized phase when $W\gtrsim W'_{2L}$. Here the independence of $f_{im}$ on the boundary condition also indicates the absence of the non-Hermitian skin effect in this case with the non-conjugate hopping phase.

\begin{figure}[tb]
	\centering
	\includegraphics[width=0.48\textwidth]{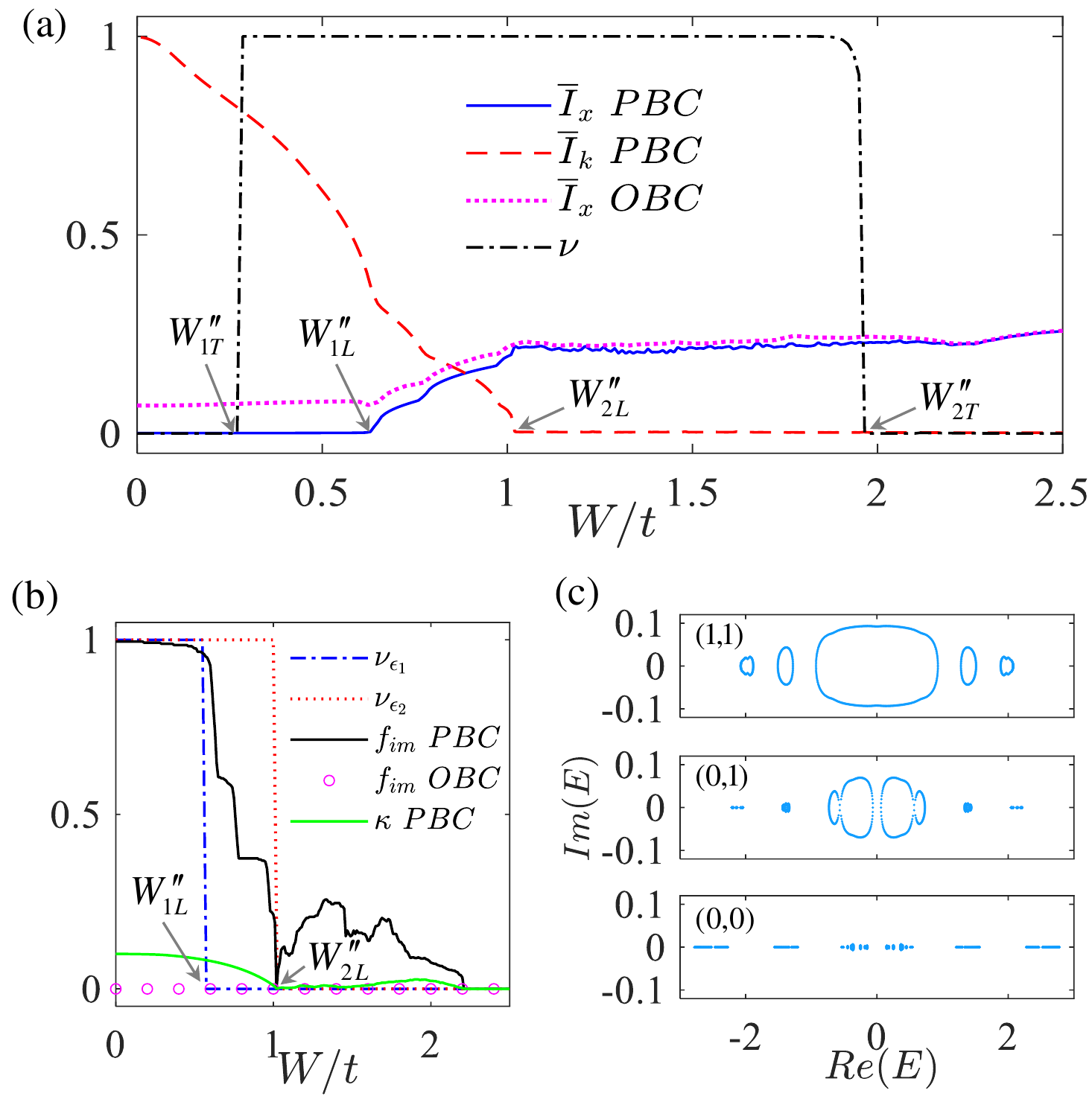}
	\caption{(Color online) (a) $\bar{I}_{x,k}$ (under the PBCs and OBCs) and $\nu$ as a function of $W$. Two topological transition points at $W''_{1T}\approx0.27$ and $W''_{2T}\approx2.0$, and two localization transition points $W''_{1L}\approx0.63$ and $W''_{2L}\approx1.02$ are labeled. (b) $\nu_{\epsilon_1}$ (blue dash-dot line), $\nu_{\epsilon_2}$ (red dot line), $f_{im}$ under PBCs (black solid line), $f_{im}$ under OBCs (pink round), maximum imaginary part of the energy spectrum $\kappa$ (green solid line) as a function of $W$. (c) Energy spectrum for $W=0.4,0.7,1.4$ (from top to bottom) with corresponding $(\nu_{\epsilon_1},\nu_{\epsilon_2})$ inside each panel. Other parameters are $m=1.02$ and $g=0.1$. From top to bottom, the energy basis are $(\epsilon_1,\epsilon_2)$=$(0,2.08)$, $(0.06,2.21)$, $(0.15,2.78)$.} \label{fig6}
\end{figure}

To characterize the topology of the energy spectrum in this non-Hermitian quasiperiodic chain, we can use the spectral winding numbers defined as \cite{ZGong2018}
\begin{equation}\label{vE}
	\nu_{\mu}=\int_{0}^{2 \pi} \frac{d \theta}{2 \pi i} \partial_{\theta} \ln \operatorname{det}\left[\mathcal{H}\left(\theta\right)-\mu \right].
\end{equation}
Here the modified Hamiltonian $\mathcal{H}(\theta)=H'(\theta)$ with respect to the periodic phase shift $\theta$ is given by rewriting $m'_n$ in Eq. (\ref{H'}) as
\begin{equation}
	m'_n(\theta)=m+W\cos(2\pi \alpha n  + \theta + ih).
\end{equation}
In addition, the indexes $\mu=\epsilon_1,\epsilon_2$ denote the real parts of eigenenergies for the eigenstates with the smallest and maximum IPRs $I_{x}^{(j)}$ for the most extended and localized eigenstates, respectively. As the eigenstate in the top (center) of the energy spectrum is the first (last) one to be localized when increasing $W$, the energy basis are given by  $\epsilon_1=\max(|\text{Re}(E_n)|)$ and $\epsilon_2=\min(|\text{Re}(E_n)|)$. Unlike the winding number $\nu$ for eigenstates, here $\nu_{\mu}$ counts how many times the eigenenergy trails enclosing the energy base $\mu$ on the complex plane when $\theta$ changes from 0 to $2\pi$. Thus, $\nu_{\mu}$ characterizes the topological structure of the complex energy spectrum, instead of the number of edge modes.

The numerical result of $\nu_{\mu}$ for $h=0.5$ as a function of $W$ is shown in Fig. \ref{fig5}(d). We find that the values of $\nu_{\epsilon_1}$ and $\nu_{\epsilon_2}$ change at $W'_{1L}$ and $W'_{2T}$, respectively. In this non-Hermitian chain, the extended and intermediate phases can be characterized by ($\nu_{\epsilon_1}$,$\nu_{\epsilon_2}$)=(0,0) and (1,0) respectively, while the localized phase takes ($\nu_{\epsilon_1}$,$\nu_{\epsilon_2}$)=(1,1) or (1,2). This correspondence can be understood in the assistance of the eigenenergies spectrum shown in Fig. \ref{fig5}(e), with the disorder strength $W=0.3,0.8,1,1.4$ (from the top to the bottom panel) for four typical cases. When $W<W'_{1L}$ with $W=0.3$, $\nu_{\epsilon_1}=\nu_{\epsilon_2}=0$ due to the real spectrum with $f_{im}=0$. When $W'_{1L}<W<W'_{2L}$ with $W=0.8$, part of eigenstates become localized with complex eigenenergies that encloses $\epsilon_1$ on the complex plane and thus $\nu_{\epsilon_1}=1$. When $W'_{2L}<W<W'_{2T}$ with $W=1$, most eigenenergies become complex with $f_{im}\approx1$ and enclose $E_{1,2}$ with $\nu_{\epsilon_1}=\nu_{\epsilon_2}=1$. Interestingly, we find that the transition from $\nu_{\epsilon_2}=1$ to $\nu_{\epsilon_2}=2$ coincides with that from $\nu=1$ to $\nu=0$ at $W'_{2T}$, due to the band crossing at this topological transition point. These results demonstrate the coincidence of disorder-induced real-complex, localization and topological transitions, as well as the existence of the three types of TAIs in this non-Hermitian quasiperiodic lattice.

\subsection{\label{IIIB} Asymmetric hopping-strength case}

We consider another kind of non-Hermiticity induced by the asymmetric hopping strength \cite{SYao2018,Kunst2018,Zhang2020,Yang2020,Okuma2020,Borgnia2020,Hatano1996,Hatano1997} in our model, which reads
\begin{equation}\label{H''}
	H''=\sum_{n=1}^{N} (m_n a_n^\dagger b_n + \text{H.c.}) + t e^{-g} a_{n+1}^\dagger b_n + t e^g b_{n}^\dagger a_{n+1}.
\end{equation}
Here $g$ acts as an imaginary gauge potential and gives rise to the asymmetric hopping, which has been experimentally realized in ultracold atoms \cite{WGou2020}, photonic systems \cite{LXiao2020}, and electrical circuits \cite{Helbig2020}. Notably, in this case $\mathcal{H}(\theta)=H''(\theta)$ in Eq. (\ref{vE}), with $H''(\theta)$ defined by adding the periodically twisted phase to the asymmetric hopping under the PBCs: $g\rightarrow g-i\theta/L$ in Eq. (\ref{H''}). We plot $\nu$ as a function of $W$ for $m=1.02$ and $g=0.1$ in Fig. \ref{fig6}(a). The disorder-induced topological transitions in this case happen at $W=W''_{1T}= W_{1T}\approx0.27$ (from trivial phase to TAIs) and $W=W''_{2T}= W_{2T}\approx2.0$ (from TAIs to trivial phase). This can be understood from the fact that $H''$ can be transformed to $H$ under OBCs through a similarity transformation $H=SH''S^{-1}$ with $S=\text{diag}\{1,1,e^{-g},e^{-g},e^{-2g},e^{-2g},...,e^{-Ng},e^{-Ng}\}$. The corresponding eigenstates $\ket{j''}$ can be obtained via $\ket{j''}=S^{-1} \ket{j}$ with the same real eigenenergies of $H$. This leads to the non-Hermitian skin effect for bulk states of $H''$ under the OBCs.

We calculate $\bar{I}_{x,k}$ and find the localization transition points $W''_{1L}\approx0.63>W_{1L}$ and $W''_{2L}=W_{2L}\approx1.02$ with increasing $g$ in Fig. \ref{fig6}(a). The asymmetry hopping tends to enlarge (keep) the extend (fully localized) phase region and reduce the intermediate phase region. Notably, the averaged IPR $\bar{I}_x$ under OBCs is larger than that under PBCs in extended and intermediate phases due to the non-Hermitian skin effect in this case \cite{DWZhang2019}. In the localized phase for $W>W''_{2L}$, $\bar{I}_x$ takes nearly the same values for OBCs and PBCs as the skin effect of bulk states is destroyed by strong disorders.

\begin{figure}[tb]
	\centering
	\includegraphics[width=0.48\textwidth]{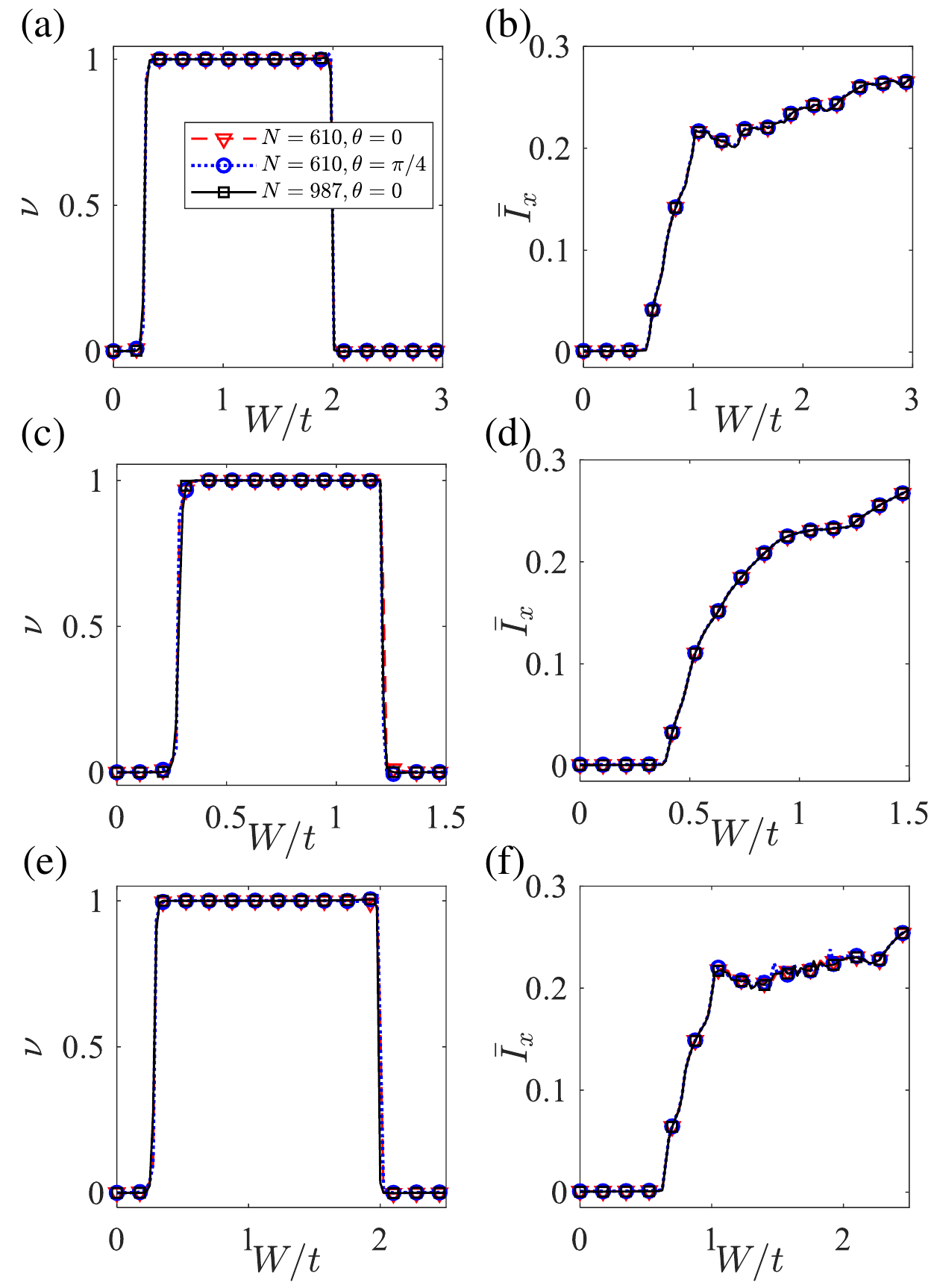}
\caption{(Color online) (a,c,e) Real-space winding number $\nu$ and (b,d,f) mean IPR $\bar{I}_x$ as functions of $W$. The results for \{$N=610,\theta=0$\}, \{$N=610,\theta=\pi/4$\}, and \{$N=987,\theta=0$\} are plotted as red dash, blue dot, and black solid lines, respectively. (a,b) correspond to the Hermitian case with $m=1.02$. (c,d) correspond to the non-conjugate hopping-phase case with $m=1.02,~h=0.5$. (e,f) correspond to the asymmetric hopping strength case with $m=1.02,~g=0.1$.} \label{fig7}
\end{figure}

In Fig. \ref{fig6}(b), we numerically calculate $\nu_{\mu}$, the imaginary fraction $f_{im}$ and the maximum imaginary part in the energy spectrum $\kappa=\max[\text{Im}(E_j)]$ as a function of $W$ for $m=1.02$ and $g=0.1$. In this asymmetric hopping case, the eigenenergies of $H''$ under OBCs are the same to those of $H$. Thus, $f_{im}=0$ for all $W$ and there is no real-complex transition under OBCs. In addition, $\nu_{\epsilon_1}$ and $\nu_{\epsilon_2}$ change at the localization transition points $W''_{1L}$ and $W''_{2L}$, respectively. However, the disorder-induced real-complex transition under PBCs happens at $W\approx2.21>W''_{2L}>W''_{1L}$. We also plot the energy spectra for $W=0.3,0.8,1.4$ in Fig. \ref{fig6}(c). When $0\leqslant W<W''_{1L}$ with $W=0.3$, complex eigenenergies enclose $\epsilon_1$ and $\epsilon_2$ with $f_{im}\approx1$ and $\nu_{\epsilon_1}=\nu_{\epsilon_2}=1$. Increasing disorder strength when $W''_{1L}<W<W''_{2L}$ with $W=0.8$, part of eigenenergies become real and thus $\nu_{\epsilon_2}=0$. When $W>W''_{2L}$ with $W=1.4$, most eigenenergies become real with $f_{im}\approx0$ and $\kappa\approx0$, such that $\nu_{\epsilon_1}=\nu_{\epsilon_2}=0$. These results demonstrate that the existence of the three types of TAIs under the asymmetric hopping. The disorder-induced localization transition coincides with the real-complex transition under PBCs, but are not related to the topological transition in this case.

From Fig. \ref{fig5}(e) and Fig. \ref{fig6}(c), we can see opposite behaviors of the complex energy winding by increasing the quasiperiodic disorder strength $W$. This is due to the fact that the disorder term is non-Hermitian in the non-conjugate hopping-phase case, while it is Hermitian in the asymmetric hopping case. In the former case, when $W=0$, the model Hamiltonian in Eq. (\ref{H'}) reduces to the Hermitian counterpart with real eigenenegies and zero energy winding number. By increasing $W$, the non-Hermiticity strength increases and a complex spectrum with nonzero $\nu_{\mu}$ exhibits after the real-complex transition. In contrast, the asymmetric hopping Hamiltonian with $g\neq0$ in Eq. (\ref{H''}) is non-Hermitian when $W=0$, which has a complex spectrum with nonzero $\nu_{\mu}$ under PBCs. Increasing the Hermitian disorder strength $W$ in this case, more and more complex eigenenergies will turn to the real counterparts, which reduces the energy winding number until $\nu_{\mu}=0$.

\section{\label{sec4}Conclusion}

Before concluding, we note our results of topological and localization properties obtained for the lattice size $N=L/2=610$ and the phase shift $\theta=0$ preserve for other $\theta$s and larger system sizes. For instance, we plot typical results of the winding number and mean IPRs for different $N$s and $\theta$s in Fig. \ref{fig7}. The results demonstrate that the system size of $N=610$ is large enough for self-averaging and neglecting the finite-size effect in the Hermitian and non-Hermitian cases.

In summary, we have explored the topology and localization of Hermitian and non-Hermitian SSH chains with quasiperiodic hopping disorders. In the Hermitian case, we have obtained the phase diagrams with topological extended, intermediate and localized phases. Due to the coexistence of topological and localization transitions, we have uncovered three types of disorder-induced TAIs with extended, intermediate, and localized bulk states in the system. We have also studied the non-Hermitian effects on the TAIs by considering two kinds of non-Hermiticities from the non-conjugate hopping phase and asymmetric hopping strength, respectively. We have shown that the three types of TAIs can preserve and exhibit some unique localization and topological properties in these non-Hermitian cases.

{\sl Note added}. After completion of this manuscript, we noticed a very recent work (arXiv:2201.00488) on a similar problem \cite{ZLu2022}, where the TAI phase with exact mobility edges was obtained in the Hermitian case. In this paper, we furthermore study the non-Hermitian TAIs.

\begin{acknowledgments}
This work was supported by the National Natural Science
Foundation of China (Grants No. 12174126 and No. 12104166), the Key-Area Research and Development Program of Guangdong Province (Grant No. 2019B030330001), the Science and Technology Program of Guangzhou (Grant No. 2019050001), and the Guangdong Basic and Applied Basic Research Foundation (Grants No. 2021A1515010315 and No. 2020A1515110290).
\end{acknowledgments}

\bibliography{reference}

\end{document}